\documentclass[twocolumn,showpacs,prb,floatfix,superscriptaddress
,cite]{revtex4}
\setcitestyle{numbers,square}
\usepackage{graphicx}
\usepackage{epstopdf}
\usepackage{color}
\usepackage{amsmath}
\usepackage{bm}

\begin{document}
\title{Binary magnetism and directional magnon transport in alkali-doped CrI$_3$}

\author{C. Bacaksiz}
\affiliation{COMMIT, Department of Physics, University of Antwerp, Groenenborgerlaan 171, B-2020 Antwerp, Belgium}

\author{M. Soenen}
\affiliation{COMMIT, Department of Physics, University of Antwerp, Groenenborgerlaan 171, B-2020 Antwerp, Belgium}

\author{D. \v{S}abani}
\affiliation{COMMIT, Department of Physics, University of Antwerp, Groenenborgerlaan 171, B-2020 Antwerp, Belgium}

\author{R. M. Menezes}
\affiliation{COMMIT, Department of Physics, University of Antwerp, Groenenborgerlaan 171, B-2020 Antwerp, Belgium}
\affiliation{Departamento de Física, Centro de Ciências Exatas e da Natureza, Universidade Federal de 
Pernambuco, Recife--PE, 50670-901, Brazil}

\author{M. V. Milo\v{s}evi\'c}
\email{milorad.milosevic@uantwerpen.be}
\affiliation{COMMIT, Department of Physics, University of Antwerp, Groenenborgerlaan 171, B-2020 Antwerp, Belgium}

\begin{abstract}
The recent realization of two-dimensional (2D) magnets, with CrI$_3$ as a pioneering example, has opened new avenues in the fields of 2D materials and magnetism. This breakthrough has been followed by extensive efforts to manipulate and exploit their magnetic properties. In this work, we investigate the adsorption of alkali-metal atoms as a route to control the magnetic behavior of monolayer CrI$_3$. We show that, upon adsorption, alkali-metal atoms donate an electron to the CrI$_3$ layer, leading to the formation of inequivalent Cr sites, with one Cr atom exhibiting an enhanced magnetic moment of $4~\mu_B$, while the other retains its typical $3~\mu_B$ moment. These modifications significantly alter the magnetic exchange interactions, including the emergence of anisotropic exchange and Dzyaloshinskii--Moriya interactions (DMI). Consequently, the doped systems exhibit non-collinear magnetic ground states, modified temperature-dependent magnetism, and anisotropic spin-wave propagation, with a preferred propagation direction that becomes increasingly pronounced with increasing dopant size. Furthermore, an asymmetry between spin-wave propagation in opposite directions is observed, giving rise to a diode-like effect, particularly for heavier dopants. This behavior is attributed to the enhanced DMI, which breaks the symmetry of the magnon dispersion. These results demonstrate that alkali-metal doping provides an effective route to tune anisotropic and nonreciprocal magnonic properties in two-dimensional magnetic materials.
\end{abstract}

\date{\today}
\pacs{
75.70.Ak,  
75.30.Gw,  
75.30.Ds,  
75.50.Dd,  
71.15.Mb   
}

\maketitle

\section{Introduction}\label{intro}

The experimental realization of intrinsic ferromagnetism in monolayer CrI$_3$~\cite{huang2017layer} marked a major milestone in condensed matter physics, demonstrating that spontaneous long-range magnetic order can persist in an atomically thin material. The stability of this two-dimensional (2D) ferromagnetic order is enabled by magnetic anisotropy, which suppresses thermal fluctuations that would otherwise destroy it according to the Mermin--Wagner theorem. This discovery, together with the observation of ferromagnetism in few-layer Cr$_2$Ge$_2$Te$_3$~\cite{gong2017discovery}, initiated extensive research on 2D van der Waals (vdW) magnets, which now include a wide range of experimentally synthesized materials~\cite{bonilla2018strong,fei2018two,kim2019micromagnetometry,wang2016raman,kim2019antiferromagnetic,gibertini2019magnetic}. These systems offer atomically sharp interfaces, gate-tunable properties, and facile integration into heterostructures, making them promising platforms for spintronic and magnonic devices as well as for exploring quantum magnetic phenomena in reduced dimensions~\cite{gibertini2019magnetic,sivadas2015magnetic,lado2017origin,xu2018interplay,vsabani2025beyond,bacaksiz2021distinctive}.

Achieving practical applications of 2D magnets requires precise control over their magnetic properties, including exchange interactions, magnetic anisotropy, and Curie temperature. Various tuning approaches have been explored, such as strain engineering~\cite{bacaksiz2021distinctive, menezes2022tailoring, zheng2018tunable, soenen2023tunable, ghojavand2024strain}, external electric fields, defect engineering, atomic substitution, and the adsorption or intercalation of foreign atoms~\cite{wang2018electric, deng2018gate}. Among these, foreign-atom adsorption is particularly attractive due to its versatility and experimental accessibility. Alkali metals (Li, Na, K, Rb) are especially suitable candidates, as they are monovalent and readily donate their valence electron to the host material, enabling controlled electron doping without severe structural disruption. This electron-donation mechanism has been explored in two closely related contexts: energy storage and magnetism.

In energy storage applications, 2D materials have attracted significant attention as electrode components in alkali-ion batteries due to their high surface area, short diffusion paths, and mechanical flexibility~\cite{rojaee2020two}. Alkali-metal intercalation can strongly modify the electronic and structural properties of layered vdW materials, a principle widely used in battery technology. More recently, this chemistry has been linked to magnetic control in vdW magnets. For example, electrochemical lithiation of CrI$_3$ has been shown to tune exchange interactions and enhance ferromagnetic ordering~\cite{wang2024room}. Similarly, Li intercalation in bilayer CrI$_3$ induces an antiferromagnetic-to-ferromagnetic transition~\cite{wu2022atomic}, while in CrSBr it leads to strongly anisotropic carrier doping~\cite{mosina2026lithium}. These findings highlight alkali-metal incorporation as an effective route to simultaneously modify electronic and magnetic properties in 2D systems.

Surface adsorption of alkali metals has also been extensively studied as a means to tune magnetism in monolayer CrI$_3$ and related materials. Previous theoretical works have shown that alkali adsorption can induce half-metallicity, enhance magnetic moments and anisotropy, and modify the Curie temperature~\cite{qin2021effects,li2021half,yang2021first,xu2020theoretical}. Experimental studies have further demonstrated that charge doping alters the magnetization, coercive field, and ordering temperature of CrI$_3$~\cite{wang2018electric}. However, the combined effect of a series of alkali dopants with increasing atomic size---which simultaneously introduces charge transfer, structural distortion, and anisotropic exchange interactions---has not been systematically explored. In particular, the impact of such doping on magnonic properties remains largely unexplored.

Spin waves (magnons) are the fundamental low-energy excitations of ordered magnets and play a central role in magnonic devices, where they act as information carriers~\cite{menezes2022tailoring, soenen2023stacking}. Asymmetric spin-wave propagation, in which magnons traveling in opposite directions exhibit different characteristics, is particularly relevant for realizing magnonic diodes and isolators. The Dzyaloshinskii--Moriya interaction (DMI) is the key mechanism responsible for breaking the symmetry of the magnon dispersion and enabling such directional behavior~\cite{di2015enhancement, zou2024dissipative}. While this effect has been studied in metallic systems and selected 2D materials, its realization through controlled chemical doping of intrinsic 2D ferromagnets remains an open question.

In this work, we address this problem by systematically investigating the adsorption of alkali-metal atoms (Li, Na, K, Rb) on monolayer CrI$_3$ using density functional theory (DFT) combined with atomistic spin dynamics (ASD) simulations. We show that electron donation from the dopants breaks the symmetry of pristine CrI$_3$ and creates a binary magnetic lattice with inequivalent Cr sites carrying 3\,$\mu_\mathrm{B}$ and 4\,$\mu_\mathrm{B}$. This symmetry breaking leads to nonequivalent exchange interactions and the emergence of significant DMI. As a result, the doped systems exhibit non-collinear magnetic ground states, modified critical temperatures, and anisotropic spin-wave propagation with a preferred direction that strengthens with increasing dopant size. Most notably, a pronounced diode-like asymmetry in magnon transmission emerges, approaching nearly complete directionality for Rb doping. These results establish alkali-metal adsorption as an effective and chemically tunable route for engineering directional magnonic functionality in 2D magnetic materials.

\section{Theoretical framework}\label{secII}

\subsection{First principles calculations}\label{compdet}

In order to investigate the structural, electronic and magnetic properties we use calculations based on density functional theory (DFT). We used the Vienna \textit{ab initio} simulation package VASP~\cite{kresse1993ab,kresse1996efficiency,kresse1996efficient} that iteratively solves the Kohn-Sham equations using a plane-wave basis set. To describe electron exchange and correlation, the Perdew-Burke-Ernzerhof (PBE) form of the generalized gradient approximation (GGA)~\cite{perdew1996generalized} was adopted. Spin-orbit coupling (SOC) was included in all calculations. The van der Waals (vdW) forces were taken into account using the DFT-D2 method of Grimme~\cite{grimme2006semiempirical}. 

The kinetic energy cut-off of the plane-wave basis set was 600 eV and energy convergence criterion was 10$^{-6}$ eV in the ground-state calculations. For large supercell calculations, the \textit{encut} value is reduced to 300 eV with the qualitative control of the results. Gaussian smearing of 0.01 eV was used and the pressures on the unit cell were decreased to a value lower than 1.0 kbar in all three directions. On-site Coulomb repulsion~\cite{dudarev1998electron} parameter, $U$, was taken as 2.65 eV for magnetic Cr atoms~\cite{sivadas2015magnetic,chittari2016electronic,jiang2019stacking}. To avoid interactions between periodically repeating monolayers in the vertical direction, our calculations were performed with a sufficiently large vacuum space of $\sim$20\AA{} between the layers. 

To obtain magnetic exchange interactions, we employ the TB2J package~\cite{he2021tb2j}, which extracts exchange parameters from a tight-binding Hamiltonian constructed using localized Wannier orbitals. The Wannier functions are generated using the Wannier90 code~\cite{pizzi2020wannier90}, interfaced with the VASP package. Heisenberg spin Hamiltonian is considered in the form:
\begin{align}\label{hamil}
 H= \sum_{i \not = j}  \big[ J_{ij}^{iso}\mathbf{S}_i \mathbf{S}_j & +   \mathbf{S}_i \mathbf{J}_{ij}^{ani} \mathbf{S}_j \nonumber 
   + \mathbf{D}_{ij}\cdot (\mathbf{S}_i \times \mathbf{S}_j) \big] \\ &+ \sum_{i} \mathbf{S}_i \mathbf{A}_{ii} \mathbf{S}_i,
\end{align}
where $\mathbf{S}_i = (S_{i}^{x},S_{i}^{y},S_{i}^{z})$ is a vector. $J_{ij}^{iso}$, $\mathbf{J}_{ij}^{ani}$, $\mathbf{D}_{ij}$ and $\mathbf{A}_{ii}$ are, respectively, the isotropic exchange parameter, the anisotropic exchange matrix, the Dzyaloshinskii-Moriya interaction (DMI) vector between the magnetic sites, and the single-ion anisotropy (SIA) matrix of $i$th site. 

The spin wave dispersion relations are calculated using \emph{SpinW}~\cite{toth2015linear}, a code based on linear spin wave theory.

\subsection{Atomistic spin dynamics}\label{ASD}

To investigate the temperature-dependent magnetization and spin-wave (SW) propagation in the considered magnetic systems, we carried out atomistic spin dynamics (ASD) simulations based on the stochastic Landau–Lifshitz–Gilbert (LLG) equation. The magnetic parameters employed in the simulations were obtained from our first-principles calculations. The simulations were performed using the package $Spirit$~\cite{muller2019spirit}, suitably modified to include the anisotropic interactions present in our Hamiltonian [Eq.~\eqref{hamil}]. The stochastic LLG equation is given by
\begin{equation}
\frac{\partial \textbf{n}_i}{\partial t}
= -\frac{\gamma}{(1+\alpha^2)\mu} \left[\textbf{n}_i \times \textbf{B}_i^\text{eff} + \alpha\textbf{n}_i \times (\textbf{n}_i \times \textbf{B}_i^\text{eff}) \right],
\end{equation}
where $\gamma$ denotes the electron gyromagnetic ratio, $\alpha$ is the Gilbert damping parameter, and $\textbf{B}_i^\text{eff} = -\partial \mathcal{H}/\partial\textbf{n}_i$ represents the effective magnetic field.

\begin{figure*}[ht]
\centering
\includegraphics[width=\linewidth]{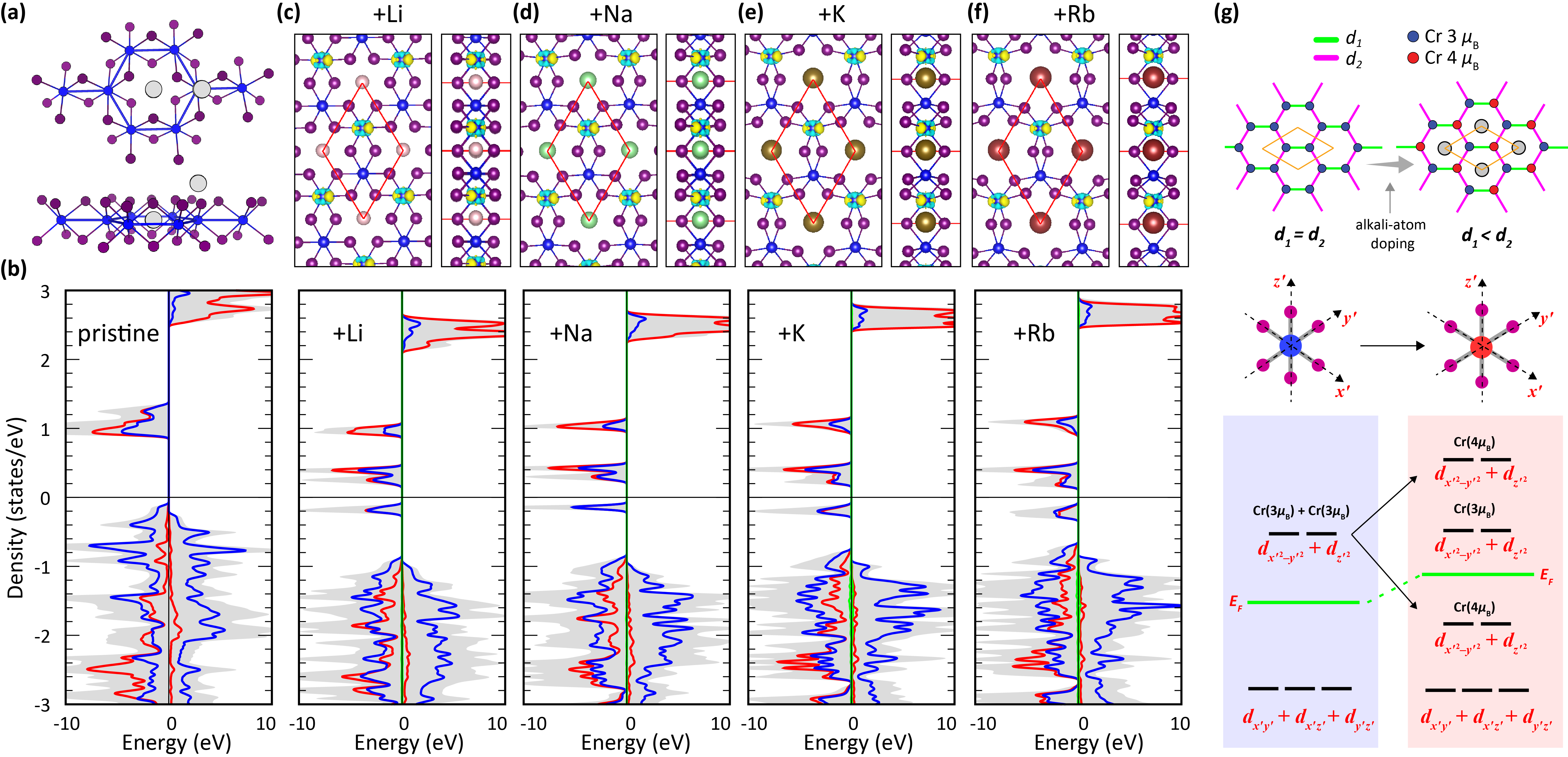}
\caption{\label{fig:st-dos} (a) Pristine CrI$_3$ and possible doping sites (gray circles), shown from the top and side views. Cr atoms are represented by blue circles. (b) Atom-projected density of states (DOS) for the pristine and doped systems (LiCrI$_3$, NaCrI$_3$, KCrI$_3$, and RbCrI$_3$). Blue, red, and green lines correspond to the Cr, I, and alkali-metal atoms, respectively, while the gray shading represents the total DOS. (c)–(f) Top and side views of the optimized structures of the doped systems, with unit cells highlighted by red parallelograms and the adatoms located at their vertices. (g) Schematic illustration of the structural distortion and the evolution of the Cr-centered octahedral coordination upon doping. The lower panel provides a simplified illustration of the splitting of the lowest-unoccupied DOS peak into three components, one of which becomes occupied.}	
\end{figure*}


For cases in which temperature effects are relevant, thermal fluctuations are incorporated through a stochastic field $\textbf{B}^\text{th}$, added to the effective field acting on each localized spin, i.e., $\textbf{B}_i^\text{eff} \rightarrow -\partial\mathcal{H}/\partial \textbf{n}_i + \textbf{B}_i^\text{th}$. The amplitude of the thermal field follows from the fluctuation–dissipation theorem and is expressed as $\textbf{B}_i^\text{th}(T,t)=\bm{\eta}_i(t)\sqrt{2\mathcal{D}/\Delta t} = \bm{\eta}_i(t)\sqrt{\frac{2\alpha k_B T}{\gamma\mu\Delta t}}$, where $T$ is the temperature, $k_B$ is the Boltzmann constant, and $\Delta t$ is the simulation time step. The term $\bm{\eta}_i(t)$ corresponds to a Gaussian white noise accounting for thermal fluctuations at each spin site $i$. The stochastic thermal field satisfies $\langle \textbf{B}_{i}^\text{th}(t)\rangle = 0$ and $\langle\textbf{B}_{ia}^\text{th}(t)\textbf{B}_{jb}^\text{th}(t')\rangle = 2\mathcal{D}\delta_{ij}\delta_{ab}\delta(t-t')$, where $a$ and $b$ denote the Cartesian components of $\textbf{B}_i^\text{th}$. 

\begin{table*}[htbp]
\caption{\label{struc_tab} Structural and electronic parameters of pristine monolayer CrI$_3$ and alkali-metal-doped systems, with a single dopant atom per unit cell. The actual formula unit, XCr$_2$I$_6$, is reduced to XCrI$_3$ (X = Li, Na, and K). Here, $a$ is the lattice constant; $d_1$ and $d_2$ denote the nearest-neighbor Cr--Cr distances along the armchair and zigzag directions, respectively; $s$ is the strain induced by the dopant atom; $\theta$ is the angle between the lattice vectors. Since the electron donated by the dopant atom is localized primarily on one of the two Cr atoms in the unit cell, the magnetic moments of the two Cr atoms are given separately as $m_1$ and $m_2$. The electronic band gap, $E_g$, and binding (adsorption) energy, $E_b$, are also shown.} 
\begin{tabular}{lcccccccccccc}
\hline\hline
                      & $a$        & $\theta$      & $d_{1}$   &   $d_{2}$            &  $s$       &   m$_{1}$    &  m$_{2}$    & $E_g$     &  $E_b$\\
                      & (\AA{})    & ($^{\circ}$) & (\AA{})    &  (\AA{})             &  ($\%$)      &   ($\mu_{B}$)    &  ($\mu_{B}$)    & (eV)        &  (eV)\\
\hline
CrI$_3$                 & 6.92      &  60.0      & 3.99      & 3.99               & $-$            & 3           &  3            &  0.84                & \\
LiCrI$_3$               & 6.98      &  61.1      &   4.02   &  4.07              & 0.8 (0.6/1.9)   & 3           &  4            &  0.13 (0.48/0.85)     & -4.17 \\
NaCrI$_3$               & 7.20      &   60.9     &   4.11   &   4.21            & 4.0 (2.9/5.3)    & 3           &  4           &  0.19 (0.52/0.93)     & -3.56 \\
KCrI$_3$                & 7.50      &  60.3      &  4.24    &  4.38              & 8.3  (6.2/9.7)  & 3           &  4           &  0.14 (0.36/0.96)     & -3.15 \\
RbCrI$_3$                & 7.61     &   58.9     & 4.29      &  4.45            & 9.9  (7.4/11.3)  & 3           &  4           &  0.10 (0.22/0.96)     & -3.00\\
\hline\hline 
\end{tabular}
\end{table*}

\section{Results and discussions}\label{results}

\subsection{Structural properties}

Pristine CrI$_3$, serving as the host material, crystallizes in the trigonal $P\Bar{3}1m$ space group. The structure consists of a hexagonal planar lattice of Cr atoms sandwiched between two triangular planar lattices of iodine atoms. Each Cr atom is octahedrally coordinated by six I atoms, while each I atom bonds to two Cr atoms. The system exhibits threefold in-plane symmetry, with the upper and lower iodine layers rotated by 180°.

When possible adsorption sites for alkali dopants are considered, two primary locations emerge: the hollow site, located at the center of the hexagonal Cr lattice, and the midpoint of the surface iodine triangles [see Fig.~\ref{fig:st-dos}~(a)]. Our calculations indicate that the hollow site is the most energetically favorable adsorption site for all considered alkali dopants. Fig.~\ref{fig:st-dos}~(c-f) shows the top and side views of the optimized
structures of the doped crystals. This doping induces an expansion of the lattice, proportional to the size of the dopant, leading to tensile strain. In Table~\ref{struc_tab}, the lattice expansion and the corresponding strain values are presented. In addition, the angle between the lattice vectors and the two nearest Cr--Cr distances, $d_1$ and $d_2$, corresponding to the armchair and zigzag directions [see Fig.~\ref{fig:st-dos}~(g)], respectively, are also listed. The perfect hexagonal symmetry is broken upon doping. In the Li-, Na-, and K-doped cases, the lattice angle becomes slightly larger than 60°, whereas in the Rb-doped case it is slightly smaller. In all doped systems, $d_1$ is shorter than $d_2$.

The adsorption energies range from $-4.17$ eV (Li) to $-3.00$ eV (Rb). Since Li is the smallest dopant, it induces the least structural distortion in CrI$_3$, resulting in a relatively stronger binding. As expected, the alkali atom donates one electron to the CrI$_3$ layer. One of the most notable results is that the donated electron localizes on only one of the two Cr atoms in the unit cell, creating a binary magnetic lattice in which one Cr atom carries $3~\mu_{B}$ and the other $4~\mu_{B}$, as illustrated in Fig.~\ref{fig:st-dos}~(g). This site-selective charge localization may provide an experimentally accessible signature of dopant-induced regions, enabling the identification of local magnetic configurations and offering a potential route for enhanced control over the magnetic properties.

\subsection{Electronic properties}

To identify the electronic modifications induced by uniform alkali-metal doping in monolayer CrI$_3$, we calculate the atomically decomposed density of states (DOS) for each structure. Since one of the two Cr atoms in the unit cell captures the electron donated by the alkali atom, we search for an electronic signature of this modification. In Fig.~\ref{fig:st-dos}~(b), the DOS of all considered structures, including pristine monolayer CrI$_3$, are shown.

The most prominent change in the DOS for all doped cases is that the lowest peak in the conduction region, originating from the $e_g$ orbitals ($d_{x^2 - y^2}$ and $d_{z^2}$) associated with the octahedral coordination shown in Fig.~\ref{fig:st-dos}(g), splits into three well-separated peaks, one of which shifts below the Fermi level, indicating that it becomes occupied. Further analysis shows that the occupied peak originates from the partially occupied $d_{x^2 - y^2}$ and $d_{z^2}$ orbitals of one of the Cr atoms in the unit cell, which consequently acquires a magnetic moment of $4~\mu_{B}$. The unoccupied portion of these orbitals of the same Cr atom appears as one of the split peaks and shifts to higher energy. The peak corresponding to the Cr atom that remains in the pristine-like configuration (with $3~\mu_{B}$) retains its general shape and energy position relative to the $t_{2g}$ orbitals, consistent with the pristine case.

\begin{figure*}[ht]
\centering
\includegraphics[width=\linewidth]{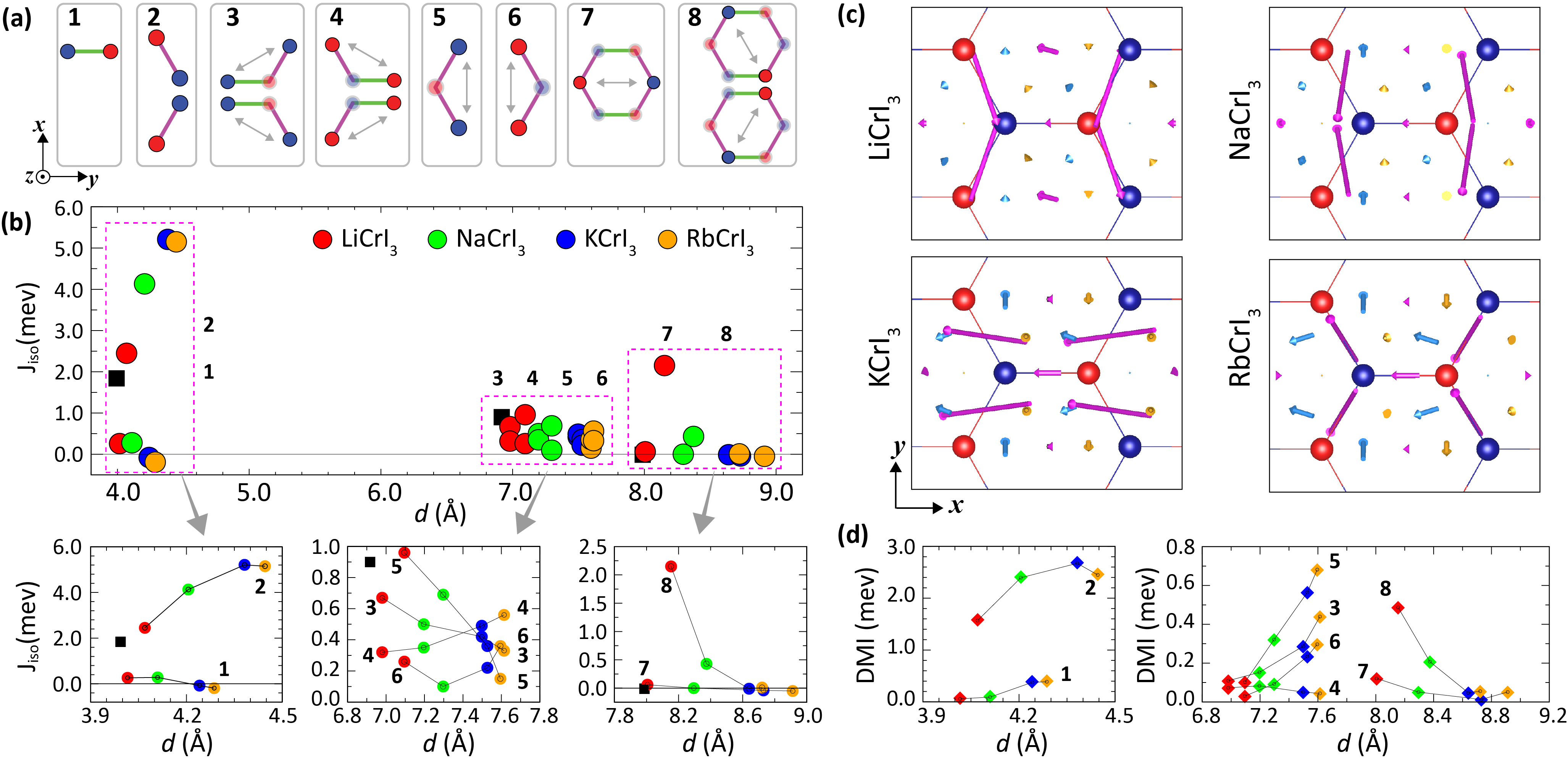}
\caption{\label{fig:exch}   (a) Classification of Cr–Cr pairs after doping. Blue and red circles represent Cr atoms with magnetic moments of $3~\mu_B$ and $4~\mu_B$, respectively. (b) Isotropic exchange interactions, $J_{ij}^{\mathrm{iso}}$, as a function of interatomic distance for all enumerated pairs shown in (a). The pristine case is indicated by black squares. Enlarged views of selected regions are provided in the lower panels. (c) Dzyaloshinskii–Moriya interaction (DMI) vectors, $\mathbf{D}_{ij}$, illustrated on the crystal structure for the different Cr–Cr bonds. Purple vectors represent the DMI for pairs 1, 2, 7 and 8; blue vectors correspond to pairs 3 and 5; and yellow vectors to pairs 4 and 6. (d) Magnitude of the DMI vectors, $|\mathbf{D}_{ij}|$, as a function of interatomic distance for all enumerated pairs.}	
\end{figure*}

\begin{figure*}[ht]
\centering
\includegraphics[width=\linewidth]{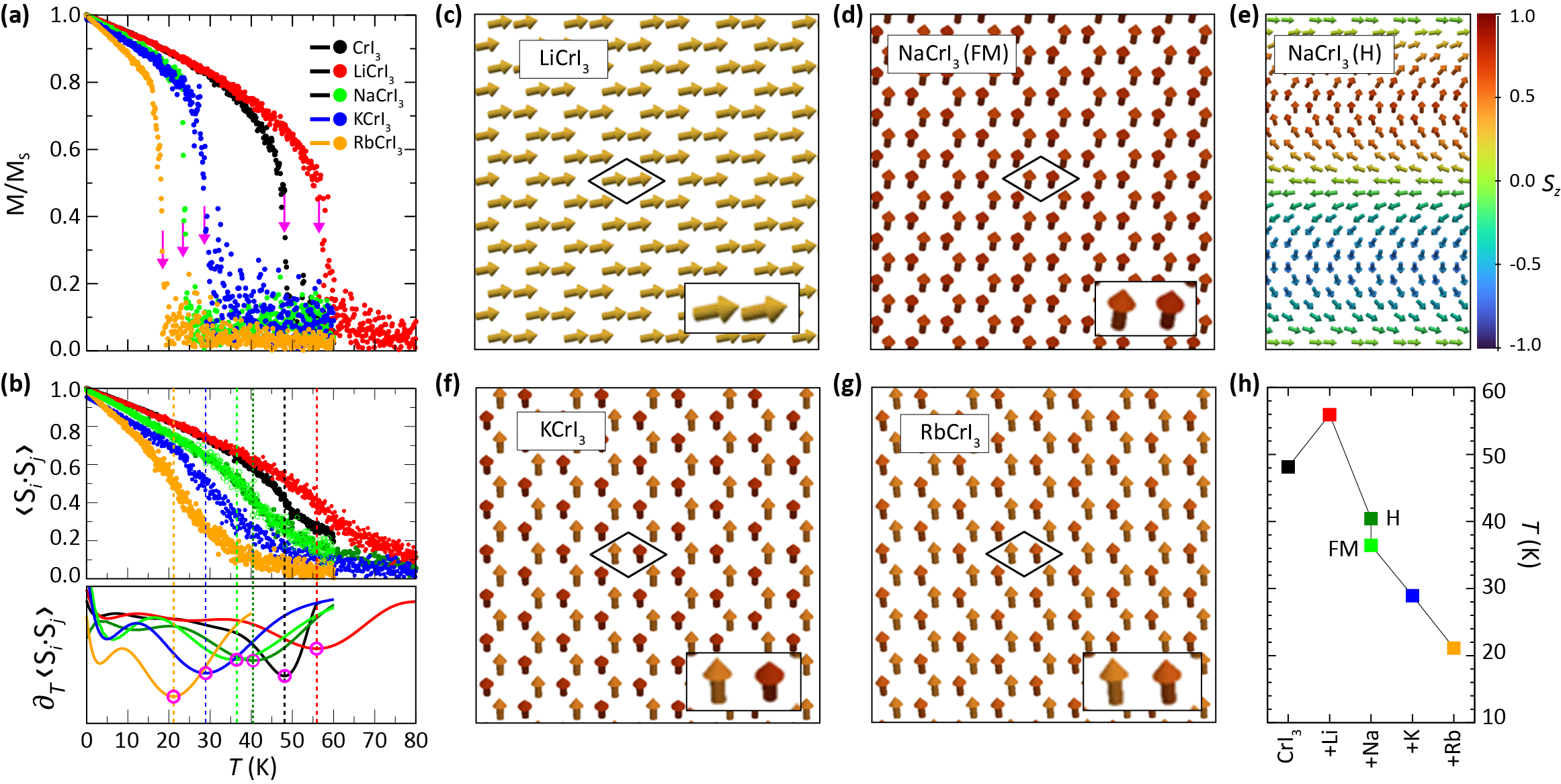}
\caption{\label{fig:tc} Temperature-dependent magnetization and ground-state (GS) magnetic order. (a) Temperature-dependent magnetization obtained from ASD simulations. (b) spin-correlation function and derivative of the spin-correlation function with respect to temperature. (c-g) Ground-state magnetic configurations obtained from ASD simulations. For NaCrI$_3$, the helical state (e) is identified as the GS, while the ferromagnetic (FM) state (d) also remains metastable. (h) Critical temperatures ($T_c$) for all considered systems, including pristine CrI$_3$.  }	
\end{figure*}

\subsection{Magnetic properties}

Modification of the electronic structure leads to significant changes in the magnetic exchange interactions and, consequently, in the Curie temperature $T_c$ and magnon properties. To quantify these effects, we extract magnetic exchange parameters within the Heisenberg model, where the interactions are represented by $3 \times 3$ exchange matrices. These matrices are decomposed into an isotropic term $J_{\mathrm{iso}}$ and an anisotropic exchange component $\mathbf{J}_{\mathrm{ani}}$, while the Dzyaloshinskii--Moriya interaction (DMI) is described by a vector $\mathbf{D}$.

As a result of the structural modifications discussed above, Cr--Cr pairs form distinct nearest-neighbor configurations. Based on these configurations, we classify and enumerate the pairs, and plot the corresponding $J_{\mathrm{iso}}$ values as a function of interatomic distance in Fig.~\ref{fig:exch}(a). The structural distortion and the emergence of binary magnetic sites lead to a splitting of the conventional neighbor shells: the first nearest neighbors (1NN) split into two distinct pair types (1 and 2), the second nearest neighbors (2NN) into four types (3--6), and the third nearest neighbors (3NN) into two types (7 and 8). In Fig.~\ref{fig:exch}(b), $J_{\mathrm{iso}}$ is shown as a function of pair distance, with enlarged views provided in the lower panels for clarity.

For pair type 1, $J_{\mathrm{iso}}$ is weakly ferromagnetic (FM) for Li- and Na-doped systems, and becomes weakly antiferromagnetic (AFM) for K- and Rb-doped cases. In contrast, pair type 2 exhibits a strong FM interaction, which increases with the size of the dopant atom. Pair types 3--6 exhibit relatively weak FM interactions for all doped systems, with $J_{\mathrm{iso}}$ values in the range of $0$--$1$~meV, showing a non-monotonic dependence on the dopant. While pair types 3 and 5 decrease with increasing dopant size, pair type 4 becomes stronger. In contrast, pair type 6 attains its lowest value in the Na-doped case and reaches its highest value for Rb doping. Pair type 7 exhibits very weak FM interactions for Li- and Na-doped systems, and very weak AFM interactions for K- and Rb-doped cases. In contrast, pair type 8 shows a relatively strong FM interaction of approximately $2$~meV in the Li-doped case, exceeding that of the first nearest neighbor (1NN) in pristine CrI$_3$. This interaction decreases rapidly with increasing dopant size, becomes weakly AFM, and eventually saturates.

The diversity in magnetic interactions originates from the asymmetry introduced by electron donation from the alkali atom, which transfers a single electron to one of the two Cr atoms in the unit cell. In addition, the varying size of the dopant modifies the crystal structure and the local coordination environment of the Cr atoms, thereby opening different magnetic exchange pathways. Such modifications also promote anisotropic magnetic interactions, particularly the Dzyaloshinskii--Moriya interaction (DMI).

In Fig.~\ref{fig:exch}(c) and (d), the DMI vectors and magnitudes between Cr atoms are illustrated, where the magnetic moments are represented by blue ($3~\mu_B$) and red ($4~\mu_B$) spheres. Similar to the behavior of $J_{\mathrm{iso}}$, pair type 2 exhibits the strongest DMI, with a significantly larger magnitude compared to the other pairs. The DMI for pair types 3, 5, and 6 increases with increasing dopant size [see Fig.~\ref{fig:exch}(d)], whereas pair types 1, 4, and 7 remain weak, and that of pair type 8 decreases.

Not only the magnitude of the DMI, but also its direction and the cumulative contribution of all DMI vectors are important for determining the magnetic ground states and spin-wave properties. For all structures, pair type 2 is the primary contributor to the directional imbalance of the DMI, and the directions of the DMI vectors are strongly affected by changes in the dopant atom, as shown in Fig.~\ref{fig:exch}(c). Moreover, in the Li-doped case, pair type 8 exhibits a finite net DMI and is expected to contribute to the magnetic ground state and spin-wave properties. Similarly, for KCrI$_3$ and RbCrI$_3$, pairs 3, 5, and 6 also make small but non-negligible contributions. In the Na-doped case, although the DMI magnitude is nonzero for all pairs, no other pair type provides a contribution comparable to that of pair type 2. This imbalance in the DMI interactions suggests that asymmetric spin-wave propagation is likely to emerge.

\subsection{Critical temperatures}

To investigate the effects of alkali-metal atom doping on the thermal properties of CrI$_3$, we perform stochastic spin dynamics simulations [see Sec.~\ref{ASD}] to obtain temperature-dependent magnetization for each considered system. For the simulations, a $24\times24$ supercell of the spin lattice was employed, with periodic boundary conditions. As a first approach, the spin system is initialized in a random configuration at high temperature and then cooled down in temperature steps of 0.125 K. To obtain the equilibrium magnetization at each temperature, the system is relaxed over $10^3$ time steps, with a time step $\Delta t = 1$ fs. 

During the cooling process, the Li-doped system evolves toward a FM configuration, where the magnetic moments are primarily aligned along the $x$ direction, with small tilts toward the $y$ and $z$ directions, as shown in Fig.~\ref{fig:tc}(c). In contrast, during the cooling process, the Na-doped system could evolve toward two different states: (\textit{i}) a helical magnetic order [see Fig.~\ref{fig:tc}(e)], in which the magnetic moments rotate in the $x$--$y$ plane while exhibit oscillations in the $z$ component, with wave vector along $y$ direction; and (\textit{ii}) a FM configuration, as shown in Fig.~\ref{fig:tc}(d). Although the FM state is formed, we found it to be a metastable configuration, and the helical state to be the ground state for Na-doped system. To determine the ideal helix pitch that minimizes the system energy, we varied the supercell size and repeated the simulations. We find that a configuration with $13$~Cr sublayers along the $y$ direction, as shown in Fig.~\ref{fig:tc}(e), yields the lowest-energy state.

For all structures, the cooling process can lead to configurations that are predominantly aligned, but may exhibit magnetic domain formation. This behavior can be attributed to the cooling protocol, in which different regions of the system independently relax into equivalent states without establishing a global alignment across the entire system. Therefore, to accurately determine the magnetic ground states, we perform additional zero-temperature simulations, in which the system is initialized in ferromagnetic configurations with different global spin orientations and subsequently relaxed at $T = 0$~K. The resulting configurations are compared based on their energies, and the lowest-energy configuration is identified as the ground state (GS).

For the K- and Rb-doped systems, the GS exhibits a nearly ferromagnetic configuration, with the magnetic moments primarily aligned along the angle bisector between the $y$ and $z$ axes, in contrast to the Li-doped case. Notably, the anisotropic interactions induce oscillations in the $z$ component of the magnetic moments between the two Cr sites, revealing a binary characteristic that emerges in these doped magnetic materials. The angle between nearest-neighbor magnetic moments is largest in the K-doped system, followed by the Rb- and Na-doped systems, as indicated by the color scale in Fig.~\ref{fig:tc}~(f), (g) and (d), respectively.

To obtain a clearer visualization of the magnetization curves while avoiding domain formation, we performed heating simulations starting from the corresponding FM states at $T=0$ and progressively increasing the temperature. Fig.~\ref{fig:tc}(a) shows the resulting magnetization curves as a function of temperature for all considered structures. Note that, for NaCrI$_3$, the magnetization exhibits a premature drop due to the formation of a helical state. Although magnetic order persists in this state, the net magnetization of the sample is suppressed.


\begin{figure*}[t]
\centering
\includegraphics[width=\linewidth]{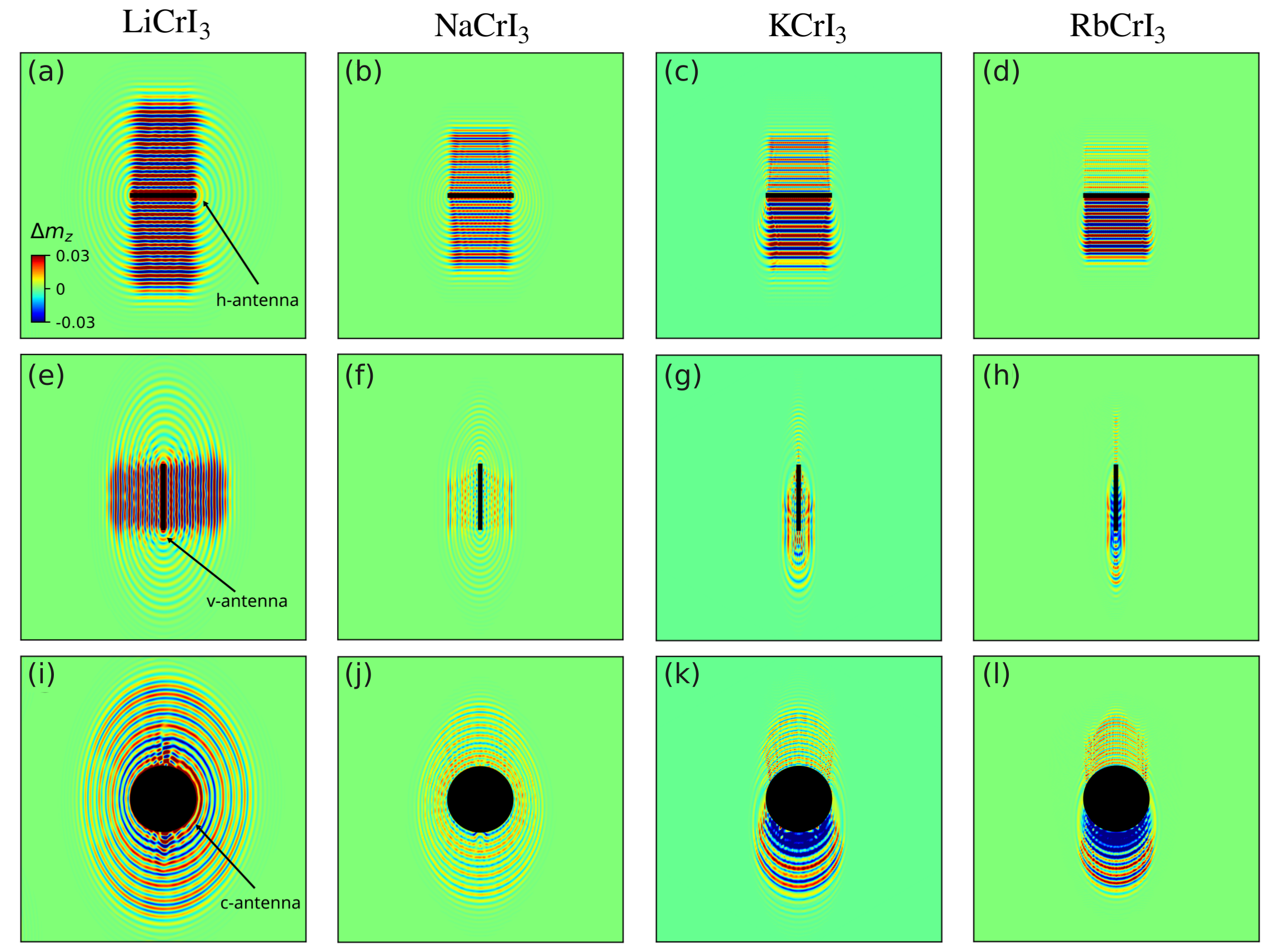}
\caption{\label{fig:SWs} Snapshots of SW propagation excited by different antenna geometries: 
(a-d) a horizontally oriented antenna (h-antenna), (e-h) a vertically oriented antenna (v-antenna), 
and (i-l) a circular antenna (c-antenna). In all cases, an oscillating magnetic field is applied 
along the $z$-direction within the antenna region. Asymmetric SW propagation is clearly 
observed across the different materials. In all cases, the excited SW frequency is $1$ THz and the simulations run for $8$ ps. }	
\end{figure*}


\begin{figure*}[ht]
\centering
\includegraphics[width=\linewidth]{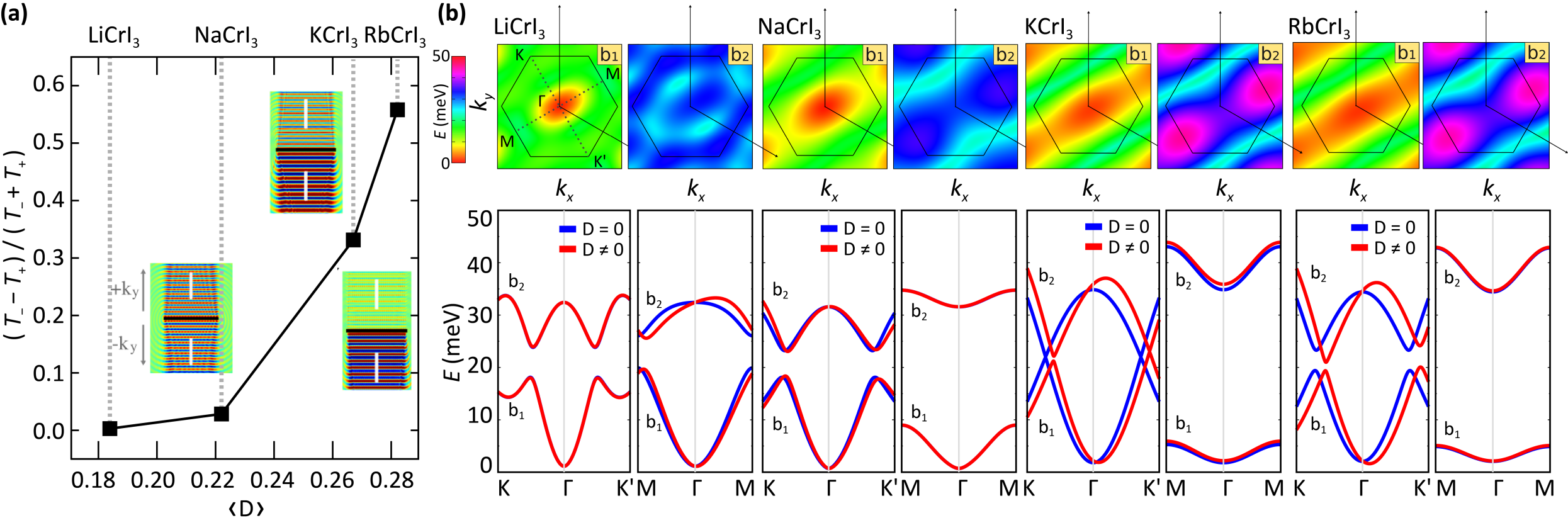}
\caption{\label{fig:diode} (a) SW transmission asymmetry as a function of the average DMI, $\langle \mathrm{D} \rangle$, for the different doped materials. Insets show examples of SW propagation, with white bars indicating the regions where the transmission amplitudes $T_{+}$ and $T_{-}$ are measured, corresponding to SW propagation along $+k_y$ and $-k_y$, respectively. In all cases, the excited SW frequency is 1 THz. (b) Magnon dispersion for each considered system. Top panels: contour plots of the lower ($b_1$) and upper ($b_2$) magnon bands. Bottom panels: magnon band structures along the K$\Gamma$K$^\prime$ and M$\Gamma$M paths in the Brillouin zone, plotted with and without DMI. These paths correspond to the SW propagation directions of the h-antenna and v-antenna in Fig.~\ref{fig:SWs}, respectively.}	
\end{figure*}

To accurately determine $T_c$ for the considered structures, we calculate the nearest-neighbor spin-correlation functions, as shown in Fig.~\ref{fig:tc}(b) for all doped structures and pristine CrI$_3$. Rather than extracting $T_c$ directly from the spin-correlation curves, we determine it from their temperature derivatives by identifying the inflection point at the highest temperature [see Fig.~\ref{fig:tc}(b), bottom panel]. For the Na-doped system, both the FM and helical states are considered, represented by the light and dark green lines in Fig.~\ref{fig:tc}(b), respectively. The resulting $T_c$ values are in very good agreement with the temperature at which the magnetization drops in Fig.~\ref{fig:tc}(a), except for the Na-doped system. In the latter, the magnetization curve exhibits a premature drop at approximately $25$~K, whereas the spin-correlation method yields $T_c \sim 36$~K and $40$~K for the FM and helical states, respectively. Importantly, magnetic order in the helical state can still be captured by the spin correlations, even though the net magnetization remains close to zero.

Figure~\ref{fig:tc}(h) summarizes the calculated $T_c$ values for all structures, including pristine CrI$_3$. The $T_c$ of the helical state is consistent with the overall trend of decreasing $T_c$ with increasing dopant size. Although all doped systems retain sizable $T_c$ values, only the Li-doped system exhibits an enhancement relative to pristine CrI$_3$, with $T_c$ increasing from $\sim 48$~K to $\sim 56$~K.

\subsection{Magnonic properties}

As alkali-metal atom doping in CrI$_3$ significantly affects the magnetic parameters and temperature-dependent magnetization, the magnonic properties of the 2D magnetic material are also expected to be modified. To investigate spin-wave (magnon) propagation in the considered doped systems, we perform spin dynamics simulations as detailed in Sec.~\ref{ASD}. For the simulations, we consider a spin lattice with $256 \times 256$ unit cells, each containing two magnetic atoms, and apply periodic boundary conditions.

For the spin-wave (SW) propagation analysis, we consider the magnetic configurations shown in Fig.~\ref{fig:tc}(c,d,f,g), where the Li-, K-, and Rb-doped systems are in their ferromagnetic ground states, while the Na-doped system is taken in its metastable FM state. These states differ significantly from pristine CrI$_3$, whose ground-state configuration consists of a saturated FM alignment along the $z$ direction. 

Fig.~\ref{fig:SWs} shows snapshots of the SW propagation for the four considered materials when excited by different antenna geometries: a horizontally oriented antenna (h-antenna) [Fig.~\ref{fig:SWs}~(a-d)], which induces wavevectors predominantly along the $y$ direction; a vertically oriented antenna (v-antenna) [Fig.~\ref{fig:SWs}~(e-h)], which induces wavevectors predominantly along the $x$ direction; and a circular antenna (c-antenna) [Fig.~\ref{fig:SWs}~(i-l)], which induces wavevectors along the radial direction. In all cases, an oscillating magnetic field is applied along the $z$ direction within the antenna region.

Anisotropic SW propagation is clearly observed across the different materials. In all cases, preferential propagation along the $y$ direction over the $x$ direction is observed, and this effect becomes more pronounced as the atomic number increases from Li$\to$Na$\to$K$\to$Rb. Such anisotropic propagation originates from the induced anisotropic exchange interaction. As shown in Fig.~\ref{fig:exch}(b), the exchange interaction along $x$-oriented bonds (bonds $1$ and $7$ in Fig.~\ref{fig:exch}) is much weaker than that along $y$-oriented bonds. This anisotropic exchange therefore favors SW propagation along the $y$ direction over the $x$ direction.

In addition, asymmetric propagation is observed along the $+y$ and $-y$ directions, and this effect also becomes more pronounced as the atomic number increases from Li$\to$Na$\to$K$\to$Rb. This asymmetric propagation results in a near diode-like effect in the case of RbCrI$_3$, where SWs predominantly propagate along the $-y$ direction, while propagation along the $+y$ direction is suppressed. Such an effect cannot be attributed to the exchange coupling $J$, since it affects the magnon dispersion symmetrically for positive and negative values of the wavevector $k_y$ [$\omega(k_y) \propto J k_y^{2}$]. The magnetic interaction capable of breaking the symmetry between positive and negative wavevectors is the DMI; therefore, we attribute the observed diode-like effect to the emergence of DMI in the doped systems.

Fig.~\ref{fig:diode}(a) shows the transmission asymmetry of SW propagation along the $y$ direction as a function of the average DMI coupling in the samples. Since we focus on propagation asymmetry along the $y$ direction, the average DMI is calculated by considering only bonds with a displacement along the $y$ direction, while bonds oriented purely along the $x$ direction, i.e., bond pairs $1$ and $7$ shown in Fig.~\ref{fig:exch}(a), are excluded. The transmission amplitudes $T_{+}$ and $T_{-}$ are calculated above and below the h-antenna, respectively, corresponding to SW propagation along $+k_y$ and $-k_y$. The white regions in the insets of Fig.~\ref{fig:diode}(a) indicate the areas where the transmission is evaluated. The transmission asymmetry clearly increases with increasing DMI, which in turn grows with the atomic number of the dopant.

For NaCrI$_3$, we also simulated SW propagation in the presence of the helical state. In this case, an additional interaction between the wavefront and the periodic domains of the helix is observed. Nevertheless, the overall anisotropic propagation persists, as in the FM case, indicating that the anisotropic propagation is governed primarily by the local magnetic interactions rather than by the stabilized magnetic configuration. 

Fig.~\ref{fig:diode}(b) shows the magnon dispersion for each doping case, where the top panels show 2D surface plots for the two characteristic bands, while the bottom panels show the dispersion transverse profiles along the K$\Gamma$K$^\prime$ and M$\Gamma$M directions, indicated by the dashed lines in the Brillouin zone. In the Li-doped system, the lower band (b$_1$) is clearly confined around the $\Gamma$ point along both the K$\Gamma$K$^\prime$ and M$\Gamma$M directions. With increasing dopant size, the low-energy region of the band extends along the M$\Gamma$M direction while remaining localized along K$\Gamma$K$^\prime$. This behavior is consistent with the directional preference of low-energy SW transport. In addition, a noticeable shift of the bands is observed, particularly for the K- and Rb-doped systems, which originates from the strong DMI. Such a shift breaks the symmetry of the magnon dispersion with respect to positive and negative wavevectors, promoting asymmetric SW propagation. This asymmetry underlies the diode-like effect observed in the simulations, especially for heavier alkali-metal dopants in CrI$_3$.

The magnon dispersion profiles along the K$\Gamma$K$^\prime$ and M$\Gamma$M directions are shown in the bottom panels of Fig.~\ref{fig:diode}(b). To highlight the role of the DMI, we additionally plot the dispersion in the absence of DMI. The effect of the DMI-induced shift differs between the lighter dopants, Li and Na, and the heavier dopants, K and Rb. In the Li- and Na-doped systems, no pronounced shift is observed around the $\Gamma$ point; however, a noticeable asymmetry appears in the higher-energy band (b$_2$), particularly for the Li case. In contrast, for the K- and Rb-doped systems, a strong shift of the dispersion is observed along the K$\Gamma$K$^\prime$ direction when DMI is included, while the overall quadratic character of the bands is preserved. Along the M$\Gamma$M direction, the dispersion becomes flatter, with no significant shift. All these features are consistent with the directional SW transport behavior discussed above.

The demonstrated ability to tune magnonic properties to achieve anisotropic and strongly asymmetric SW propagation, including diode-like behavior, establishes alkali-metal doping of 2D magnetic materials as a promising strategy for engineering magnonic and spintronic devices with intrinsic directional functionality.


\section{Conclusion}\label{conc}
In summary, we have systematically investigated the effects of alkali-metal atom adsorption (Li, Na, K, and Rb) on the structural, electronic, and magnetic properties of monolayer CrI$_3$ by combining first-principles calculations with atomistic spin dynamics simulations. We showed that electron donation from the dopant breaks the symmetry of pristine CrI$_3$, leading to the formation of nonequivalent Cr sites with magnetic moments of either $3~\mu_B$ or $4~\mu_B$. This charge-induced symmetry breaking gives rise to anisotropic exchange interactions and sizable Dzyaloshinskii--Moriya interactions (DMI), which are absent in the pristine system.

As a result, both the magnetic ground states and their thermal stability are significantly modified. While Li-, K-, and Rb-doped systems stabilize ferromagnetic ground states, the Na-doped system exhibits a helical magnetic ground state with a metastable ferromagnetic configuration. The Curie temperature $T_c$ varies systematically with dopant size, showing an enhancement for Li doping and a reduction for heavier dopants.

Furthermore, the induced anisotropic exchange and DMI lead to pronounced anisotropic and asymmetric spin-wave propagation. In all doped systems, spin waves preferentially propagate along a specific crystallographic direction, with the anisotropy becoming stronger for larger dopants. In addition, a clear asymmetry between propagation along opposite directions is observed, resulting in diode-like behavior that is particularly pronounced for Rb doping. This nonreciprocal magnon transport originates from the strength and directional imbalance of the DMI.

Our results demonstrate that alkali-metal adsorption provides a simple and effective route to engineer anisotropic and nonreciprocal magnonic behavior in two-dimensional magnetic materials. These findings open promising avenues for designing chemically-tunable magnonic and spintronic devices with intrinsic directional functionality, such as magnonic diodes and logic elements.

\begin{acknowledgments}
This work was supported by the Research Foundation - Flanders (FWO). C.B. is a senior post–doctoral fellow of FWO under contract No. 12E8823N. M.S. is an FWO doctoral fellow under grant No. 11O1423N. R.M.M. acknowledges the INCT project Advanced Quantum Materials (Proc. 408766/2024-7), involving the Brazilian agencies CNPq, FAPESP, and CAPES. The computational resources used in this work were provided by the VSC (Flemish Supercomputer Center), funded by Research Foundation-Flanders (FWO) and the Flemish Government -- department EWI.
\end{acknowledgments}

\bibliography{bib_file}

@article{huang2017layer,
  title={Layer-dependent ferromagnetism in a van der Waals crystal down to the monolayer limit},
  author={Huang, Bevin and Clark, Genevieve and Navarro-Moratalla, Efr{\'e}n and Klein, Dahlia R and Cheng, Ran and Seyler, Kyle L and Zhong, Ding and Schmidgall, Emma and McGuire, Michael A and Cobden, David H and others},
  journal={Nature},
  volume={546},
  number={7657},
  pages={270--273},
  year={2017},
  publisher={Nature Publishing Group UK London}
}

@article{gong2017discovery,
  title={Discovery of intrinsic ferromagnetism in two-dimensional van der Waals crystals},
  author={Gong, Cheng and Li, Lin and Li, Zhenglu and Ji, Huiwen and Stern, Alex and Xia, Yang and Cao, Ting and Bao, Wei and Wang, Chenzhe and Wang, Yuan and others},
  journal={Nature},
  volume={546},
  number={7657},
  pages={265--269},
  year={2017},
  publisher={Nature Publishing Group UK London}
}

@article{gibertini2019magnetic,
  title={Magnetic 2D materials and heterostructures},
  author={Gibertini, Magnetic and Koperski, Maciej and Morpurgo, Alberto F and Novoselov, Konstantin S},
  journal={Nature nanotechnology},
  volume={14},
  number={5},
  pages={408--419},
  year={2019},
  publisher={Nature Publishing Group UK London}
}

@article{bonilla2018strong,
  title={Strong room-temperature ferromagnetism in VSe2 monolayers on van der Waals substrates},
  author={Bonilla, Manuel and Kolekar, Sadhu and Ma, Yujing and Diaz, Horacio Coy and Kalappattil, Vijaysankar and Das, Raja and Eggers, Tatiana and Gutierrez, Humberto R and Phan, Manh-Huong and Batzill, Matthias},
  journal={Nature nanotechnology},
  volume={13},
  number={4},
  pages={289--293},
  year={2018},
  publisher={Nature Publishing Group UK London}
}

@article{fei2018two,
  title={Two-dimensional itinerant ferromagnetism in atomically thin Fe3GeTe2},
  author={Fei, Zaiyao and Huang, Bevin and Malinowski, Paul and Wang, Wenbo and Song, Tiancheng and Sanchez, Joshua and Yao, Wang and Xiao, Di and Zhu, Xiaoyang and May, Andrew F and others},
  journal={Nature materials},
  volume={17},
  number={9},
  pages={778--782},
  year={2018},
  publisher={Nature Publishing Group UK London}
}

@article{kim2019micromagnetometry,
  title={Micromagnetometry of two-dimensional ferromagnets},
  author={Kim, Minsoo and Kumaravadivel, Piranavan and Birkbeck, John and Kuang, Wenjun and Xu, Shuigang G and Hopkinson, DG and Knolle, Johannes and McClarty, Paul A and Berdyugin, AI and Ben Shalom, M and others},
  journal={Nature Electronics},
  volume={2},
  number={10},
  pages={457--463},
  year={2019},
  publisher={Nature Publishing Group UK London}
}

@article{wang2016raman,
  title={Raman spectroscopy of atomically thin two-dimensional magnetic iron phosphorus trisulfide (FePS3) crystals},
  author={Wang, Xingzhi and Du, Kezhao and Fredrik Liu, Yu Yang and Hu, Peng and Zhang, Jun and Zhang, Qing and Owen, Man Hon Samuel and Lu, Xin and Gan, Chee Kwan and Sengupta, Pinaki and others},
  journal={2D Materials},
  volume={3},
  number={3},
  pages={031009},
  year={2016},
  publisher={IOP Publishing}
}

@article{kim2019antiferromagnetic,
  title={Antiferromagnetic ordering in van der Waals 2D magnetic material MnPS3 probed by Raman spectroscopy},
  author={Kim, Kangwon and Lim, Soo Yeon and Kim, Jungcheol and Lee, Jae-Ung and Lee, Sungmin and Kim, Pilkwang and Park, Kisoo and Son, Suhan and Park, Cheol-Hwan and Park, Je-Geun and others},
  journal={2D Materials},
  volume={6},
  number={4},
  pages={041001},
  year={2019},
  publisher={IOP Publishing}
}

@article{sivadas2015magnetic,
  title={Magnetic ground state of semiconducting transition-metal trichalcogenide monolayers},
  author={Sivadas, Nikhil and Daniels, Matthew W and Swendsen, Robert H and Okamoto, Satoshi and Xiao, Di},
  journal={Physical Review B},
  volume={91},
  number={23},
  pages={235425},
  year={2015},
  publisher={APS}
}

@article{lado2017origin,
  title={On the origin of magnetic anisotropy in two dimensional CrI3},
  author={Lado, Jose L and Fern{\'a}ndez-Rossier, Joaqu{\'\i}n},
  journal={2D Materials},
  volume={4},
  number={3},
  pages={035002},
  year={2017},
  publisher={IOP Publishing}
}

@article{xu2018interplay,
  title={Interplay between Kitaev interaction and single ion anisotropy in ferromagnetic CrI3 and CrGeTe3 monolayers},
  author={Xu, Changsong and Feng, Junsheng and Xiang, Hongjun and Bellaiche, Laurent},
  journal={npj Computational Materials},
  volume={4},
  number={1},
  pages={57},
  year={2018},
  publisher={Nature Publishing Group UK London}
}

@article{bacaksiz2021distinctive,
  title={Distinctive magnetic properties of Cr I 3 and Cr Br 3 monolayers caused by spin-orbit coupling},
  author={Bacaksiz, C and {\v{S}}abani, D and Menezes, RM and Milo{\v{s}}evi{\'c}, MV},
  journal={Physical Review B},
  volume={103},
  number={12},
  pages={125418},
  year={2021},
  publisher={APS}
}

@article{vsabani2025beyond,
  title={Beyond orbitally resolved magnetic exchange in CrI 3 and NiI 2},
  author={{\v{S}}abani, Denis and Bacaksiz, C and Milo{\v{s}}evi{\'c}, Milorad V},
  journal={Physical Review Letters},
  volume={135},
  number={3},
  pages={036704},
  year={2025},
  publisher={APS}
}

@article{menezes2022tailoring,
  title={Tailoring high-frequency magnonics in monolayer chromium trihalides},
  author={Menezes, Ra{\'\i} M and {\v{S}}abani, Denis and Bacaksiz, Cihan and de Souza Silva, Cl{\'e}cio C and Milo{\v{s}}evi{\'c}, Milorad V},
  journal={2D Materials},
  volume={9},
  number={2},
  pages={025021},
  year={2022},
  publisher={IOP Publishing}
}

@article{zheng2018tunable,
  title={Tunable spin states in the two-dimensional magnet CrI3},
  author={Zheng, Fawei and Zhao, Jize and Liu, Zheng and Li, Menglei and Zhou, Mei and Zhang, Shengbai and Zhang, Ping},
  journal={Nanoscale},
  volume={10},
  number={29},
  pages={14298--14303},
  year={2018},
  publisher={The Royal Society of Chemistry}
}

@article{soenen2023tunable,
  title={Tunable magnon topology in monolayer CrI 3 under external stimuli},
  author={Soenen, Maarten and Milo{\v{s}}evi{\'c}, Milorad V},
  journal={Physical Review Materials},
  volume={7},
  number={8},
  pages={084402},
  year={2023},
  publisher={APS}
}

@article{ghojavand2024strain,
  title={Strain-tunable magnetic and magnonic states in Ni-dihalide monolayers},
  author={Ghojavand, Ali and Soenen, Maarten and Rezaei, Nafise and Alaei, Mojtaba and Sevik, Cem and Milo{\v{s}}evi{\'c}, Milorad V},
  journal={Physical Review Materials},
  volume={8},
  number={11},
  pages={114401},
  year={2024},
  publisher={APS}
}

@article{wang2018electric,
  title={Electric-field control of magnetism in a few-layered van der Waals ferromagnetic semiconductor},
  author={Wang, Zhi and Zhang, Tongyao and Ding, Mei and Dong, Baojuan and Li, Yanxu and Chen, Maolin and Li, Xiaoxi and Huang, Jianqi and Wang, Hanwen and Zhao, Xiaotian and others},
  journal={Nature nanotechnology},
  volume={13},
  number={7},
  pages={554--559},
  year={2018},
  publisher={Nature Publishing Group UK London}
}

@article{deng2018gate,
  title={Gate-tunable room-temperature ferromagnetism in two-dimensional Fe3GeTe2},
  author={Deng, Yujun and Yu, Yijun and Song, Yichen and Zhang, Jingzhao and Wang, Nai Zhou and Sun, Zeyuan and Yi, Yangfan and Wu, Yi Zheng and Wu, Shiwei and Zhu, Junyi and others},
  journal={Nature},
  volume={563},
  number={7729},
  pages={94--99},
  year={2018},
  publisher={Nature Publishing Group UK London}
}

@article{rojaee2020two,
  title={Two-dimensional materials to address the lithium battery challenges},
  author={Rojaee, Ramin and Shahbazian-Yassar, Reza},
  journal={ACS nano},
  volume={14},
  number={3},
  pages={2628--2658},
  year={2020},
  publisher={ACS Publications}
}

@article{wang2024room,
  title={Room-temperature CrI3 magnets through lithiation},
  author={Wang, Zhongxuan and Zheng, Huafei and Chen, Amy and Ma, Lei and Hong, Stephanie J and Rodriguez, Efrain E and Woehl, Taylor J and Shi, Su-Fei and Parker, Thomas and Ren, Shenqiang},
  journal={ACS nano},
  volume={18},
  number={34},
  pages={23058--23066},
  year={2024},
  publisher={ACS Publications}
}

@article{wu2022atomic,
  title={Atomic intercalation induced spin-flip transition in bilayer CrI3},
  author={Wu, Dongsi and Zhao, Ying and Yang, Yibin and Huang, Le and Xiao, Ye and Chen, Shanshan and Zhao, Yu},
  journal={Nanomaterials},
  volume={12},
  number={9},
  pages={1420},
  year={2022},
  publisher={MDPI}
}

@article{mosina2026lithium,
  title={Lithium Intercalation in the Anisotropic Van Der Waals Semiconductor CrSBr},
  author={Mosina, Kseniia and S{\"o}ll, Aljoscha and {\v{S}}turala, Ji{\v{r}}{\'\i} and Vesel{\`y}, Martin and Levinsky, Petr and Kaman, Ond{\v{r}}ej and Dirnberger, Florian and Materzanini, Giuliana and Marzari, Nicola and Rignanese, Gian-Marco and others},
  journal={Advanced Functional Materials},
  volume={36},
  number={34},
  pages={e23178},
  year={2026},
  publisher={Wiley Online Library}
}

@article{qin2021effects,
  title={Effects of Li adsorption on the physical properties of CrBr3 and CrCl3: From monolayer to multilayer},
  author={Qin, Guojun and Li, Shujing and Shao, Xiaohong},
  journal={AIP Advances},
  volume={11},
  number={8},
  year={2021},
  publisher={AIP Publishing}
}

@article{li2021half,
  title={Half-metallicity and enhanced magnetism in monolayer T-CrTe2 by lithium adsorption},
  author={Li, Aolin and Zhou, Wenzhe and Peng, Shenglin and Wang, Yunpeng and Long, Mengqiu and Ouyang, Fangping},
  journal={Physics Letters A},
  volume={394},
  pages={127195},
  year={2021},
  publisher={Elsevier}
}

@article{yang2021first,
  title={First-principles study on the electronic properties and enhanced ferromagnetism of alkali metals adsorbed monolayer CrI3},
  author={Yang, Yidong and Li, Pengfei and Wang, Wennan and Zhang, Xianbin},
  journal={Vacuum},
  volume={194},
  pages={110561},
  year={2021},
  publisher={Elsevier}
}

@article{xu2020theoretical,
  title={Theoretical study of enhanced ferromagnetism and tunable magnetic anisotropy of monolayer CrI3 by surface adsorption},
  author={Xu, Qin-Fang and Xie, Wen-Qiang and Lu, Zhi-Wei and Zhao, Yu-Jun},
  journal={Physics Letters A},
  volume={384},
  number={29},
  pages={126754},
  year={2020},
  publisher={Elsevier}
}

@article{soenen2023stacking,
  title={Stacking-dependent topological magnons in bilayer CrI 3},
  author={Soenen, Maarten and Bacaksiz, Cihan and Menezes, Ra{\'\i} M and Milo{\v{s}}evi{\'c}, Milorad V},
  journal={Physical Review Materials},
  volume={7},
  number={2},
  pages={024421},
  year={2023},
  publisher={APS}
}

@article{di2015enhancement,
  title={Enhancement of spin-wave nonreciprocity in magnonic crystals via synthetic antiferromagnetic coupling},
  author={Di, Kai and Feng, SX and Piramanayagam, SN and Zhang, VL and Lim, Hock Siah and Ng, Ser Choon and Kuok, Meng Hau},
  journal={Scientific reports},
  volume={5},
  number={1},
  pages={10153},
  year={2015},
  publisher={Nature Publishing Group UK London}
}

@article{zou2024dissipative,
  title={Dissipative spin-wave diode and nonreciprocal magnonic amplifier},
  author={Zou, Ji and Bosco, Stefano and Thingstad, Even and Klinovaja, Jelena and Loss, Daniel},
  journal={Physical Review Letters},
  volume={132},
  number={3},
  pages={036701},
  year={2024},
  publisher={APS}
}

@article{kresse1993ab,
  title={Ab initio molecular dynamics for liquid metals},
  author={Kresse, Georg and Hafner, J{\"u}rgen},
  journal={Physical review B},
  volume={47},
  number={1},
  pages={558},
  year={1993},
  publisher={APS}
}

@article{kresse1996efficiency,
  title={Efficiency of ab-initio total energy calculations for metals and semiconductors using a plane-wave basis set},
  author={Kresse, Georg and Furthm{\"u}ller, J{\"u}rgen},
  journal={Computational materials science},
  volume={6},
  number={1},
  pages={15--50},
  year={1996},
  publisher={Elsevier}
}

@article{kresse1996efficient,
  title={Efficient iterative schemes for ab initio total-energy calculations using a plane-wave basis set},
  author={Kresse, Georg and Furthm{\"u}ller, J{\"u}rgen},
  journal={Physical review B},
  volume={54},
  number={16},
  pages={11169},
  year={1996},
  publisher={APS}
}

@article{perdew1996generalized,
  title={Generalized gradient approximation made simple},
  author={Perdew, John P and Burke, Kieron and Ernzerhof, Matthias},
  journal={Physical review letters},
  volume={77},
  number={18},
  pages={3865},
  year={1996},
  publisher={APS}
}

@article{grimme2006semiempirical,
  title={Semiempirical GGA-type density functional constructed with a long-range dispersion correction},
  author={Grimme, Stefan},
  journal={Journal of computational chemistry},
  volume={27},
  number={15},
  pages={1787--1799},
  year={2006},
  publisher={Wiley Online Library}
}

@article{dudarev1998electron,
  title={Electron-energy-loss spectra and the structural stability of nickel oxide: An LSDA+ U study},
  author={Dudarev, Sergei L and Botton, Gianluigi A and Savrasov, Sergey Y and Humphreys, Colin J and Sutton, Adrian P},
  journal={Physical Review B},
  volume={57},
  number={3},
  pages={1505},
  year={1998},
  publisher={APS}
}

@article{chittari2016electronic,
  title={Electronic and magnetic properties of single-layer MPX 3 metal phosphorous trichalcogenides},
  author={Chittari, Bheema Lingam and Park, Youngju and Lee, Dongkyu and Han, Moonsup and MacDonald, Allan H and Hwang, Euyheon and Jung, Jeil},
  journal={Physical Review B},
  volume={94},
  number={18},
  pages={184428},
  year={2016},
  publisher={APS}
}

@article{jiang2019stacking,
  title={Stacking tunable interlayer magnetism in bilayer CrI 3},
  author={Jiang, Peiheng and Wang, Cong and Chen, Dachuan and Zhong, Zhicheng and Yuan, Zhe and Lu, Zhong-Yi and Ji, Wei},
  journal={Physical Review B},
  volume={99},
  number={14},
  pages={144401},
  year={2019},
  publisher={APS}
}

@article{he2021tb2j,
  title={TB2J: A python package for computing magnetic interaction parameters},
  author={He, Xu and Helbig, Nicole and Verstraete, Matthieu J and Bousquet, Eric},
  journal={Computer Physics Communications},
  volume={264},
  pages={107938},
  year={2021},
  publisher={Elsevier}
}

@article{pizzi2020wannier90,
  title={Wannier90 as a community code: new features and applications},
  author={Pizzi, Giovanni and Vitale, Valerio and Arita, Ryotaro and Bl{\"u}gel, Stefan and Freimuth, Frank and G{\'e}ranton, Guillaume and Gibertini, Marco and Gresch, Dominik and Johnson, Charles and Koretsune, Takashi and others},
  journal={Journal of Physics: Condensed Matter},
  volume={32},
  number={16},
  pages={165902},
  year={2020},
  publisher={IOP Publishing}
}

@article{toth2015linear,
  title={Linear spin wave theory for single-Q incommensurate magnetic structures},
  author={Toth, S and Lake, B},
  journal={Journal of Physics: Condensed Matter},
  volume={27},
  number={16},
  pages={166002},
  year={2015},
  publisher={IOP Publishing}
}

@article{muller2019spirit,
  title={Spirit: Multifunctional framework for atomistic spin simulations},
  author={M{\"u}ller, Gideon P and Hoffmann, Markus and Di{\ss}elkamp, Constantin and Sch{\"u}rhoff, Daniel and Mavros, Stefanos and Sallermann, Moritz and Kiselev, Nikolai S and J{\'o}nsson, Hannes and Bl{\"u}gel, Stefan},
  journal={Physical review b},
  volume={99},
  number={22},
  pages={224414},
  year={2019},
  publisher={APS}
}

\end{document}